\documentclass[12pt]{article}
\usepackage[dvips]{graphicx}
\usepackage{feynmf}
\usepackage{amsmath,amssymb}
\usepackage{epsfig}
\usepackage{subfigure}
\usepackage{amsmath}
\usepackage{amsfonts}
\usepackage{amssymb}
\usepackage{url}
\usepackage{hyperref}
\newcommand{\be}{\begin{equation}}
\newcommand{\ee}{\end{equation}}
\newcommand{\bea}{\begin{eqnarray}}
\newcommand{\eea}{\end{eqnarray}}

\begin{document}
\begin{titlepage}
\begin{flushright}
\end{flushright}
\vspace{4\baselineskip}
\begin{center}
{\Large\bf 
 Realistic Hybrid Inflation in 5D Orbifold SO(10) GUT          
}
\end{center}
\vspace{1cm}
\begin{center}
{\large Takeshi Fukuyama $^{a,b,}$
\footnote{\tt E-mail:fukuyama@se.ritsumei.ac.jp},
Nobuchika Okada $^{c,d,}$
\footnote{\tt E-mail:okadan@post.kek.jp} and 
Toshiyuki Osaka $^{a,}$
\footnote{\tt E-mail:rp006002@se.ritsumei.ac.jp}}
\end{center}
\vspace{0.2cm}
\begin{center}
${}^{a}$ {\small \it Department of Physics, Ritsumeikan University,
 Kusatsu, Shiga, 525-8577, Japan}\\
${}^{b}$ {\small \it Ritsumeikan Global Innovation Research 
 Organization, Ritsumeikan University, \\ 
 Kusatsu, Shiga, 525-8577, Japan}\\ 
${}^{c}$ {\small \it Department of Physics,
 University of Maryland, College Park, MD 20742, USA}\\ 
${}^{d}$ {\small \it Theory Division, KEK,
Oho 1-1, Tsukuba, Ibaraki, 305-0801, Japan}\\
\medskip
\vskip 10mm
\end{center}
\vskip 10mm

\begin{abstract}
We discuss the smooth hybrid inflation scenario 
 in the context of a simple supersymmetric SO(10) GUT in 5D orbifold. 
In this GUT model, the SO(10) gauge symmetry is broken down to 
 the Pati-Salam (PS) gauge group, 
 SU(4)$_c \times$ SU(2)$_L \times$ SU(2)$_R$, 
 by orbifold boundary conditions 
 and all matter and Higgs multiplets are placed only 
 on the brane (PS brane) where only the PS symmetry is manifest. 
Further breaking of the Pati-Salam group to the Standard Model 
 one is realized by VEVs of the Higgs multiplets 
 $({\bf 4},{\bf 1},{\bf 2}) \oplus (\overline{{\bf 4}},{\bf 1},{\bf 2})$. 
The gauge coupling unification is successfully realized 
 at $M_{\rm GUT} =4.6 \times 10^{17}$ GeV after incorporating 
 the threshold corrections of the Kaluza-Klein modes, 
 with the compactification scale (assumed to be the same as 
 the PS symmetry breaking scale) 
 $M_c = v_{\rm PS}= 1.2 \times 10^{16}$ GeV. 
We show that this orbifold GUT model can naturally leads us to 
 the smooth hybrid inflation, which tunes out to be consistent with 
 the WMAP 5-year data with the predicted $M_{\rm GUT}$ and $v_{PS}$ 
 in the model. 

\end{abstract}
\end{titlepage}

\section{Introduction}
In these decades, terrestrial observational data 
 on neutrino oscillations, B physics as well as 
 the astrophysical ones like Wilkinson Microwave Anisotropy Probe
 (WMAP) give the very important informations on physics 
 beyond the Standard Model (SM), especially on 
 the grand unified models (GUTs). 
Among several GUTs, the model based on the gauge group SO(10) 
 is particularly attractive. 
In fact, SO(10) is the smallest simple gauge group 
 under which the entire SM matter contents 
 of each generation are unified into a single anomaly-free 
 irreducible representation, ${\bf 16}$. 
In these SO(10) GUTs, the so-called renormalizable minimal SO(10) model 
 (hereafter the minimal SO(10) GUT) has been paid a particular attention, 
 where two Higgs multiplets $\{{\bf 10} \oplus {\bf \overline{126}}\}$ 
 are utilized for the Yukawa couplings with matters 
 ${\bf 16}_i$ ($i=1,2,3$ is the generation index) 
 \cite{Babu} \cite{Fukuyama1} \cite{Fukuyama2}. 
A remarkable feature of the model is its high predictive power 
 for the neutrino oscillation parameters with reproducing 
 charged fermion masses and mixing angles. 
However, after KamLAND data \cite{Eguchi:2002dm} was released, 
 it entered to the stage of precision measurements, and many authors 
 performed data-fitting analysis to match up these new data.

Also, Higgs superpotential in the minimal SO(10) GUT has been 
 constructed and the detailed analysis of symmetry breaking patterns 
 have been extensively studied \cite{Fukuyama:2004xs, Bajc:2004xe}. 
This construction gives the vacuum expectation values (VEVs)
 at intermediate energy scales unlike normal SUSY GUTs, 
 which gives rise to a trouble in the gauge coupling unification 
 as well as the necessary scales of the seesaw mechanism 
 and leptogenesis. 
This mismatch of gauge couplings has been explicitly shown 
 in Ref.~\cite{Bertolini}, where they are not unified any more 
 and even the SU(2) gauge coupling blows up far below the GUT scale.

In addition to the issue of the gauge coupling unification, 
 the minimal SO(10) model potentially suffers from the problem 
 that the gauge coupling blows up around the GUT scale. 
This is because the model includes many Higgs multiplets of 
 higher dimensional representations. 
In field theoretical point of view, this fact implies 
 that the GUT scale is a cutoff scale of the model, 
 and more fundamental description of the minimal SO(10) model 
 would exist above the GUT scale.

In order to solve these problems, the minimal SO(10) GUT 
 has been considered in 5D \cite{F-K-O} with the warped background 
 geometry \cite{RS}, where the GUT gauge symmetry is assumed 
 to be broken by VEVs of Higgs multiplets on a brane, 
 as usual in 4D models.  
Another possibility of constructing GUT models in extra-dimensions 
 is to consider the so-called orbifold GUT \cite{Kawamura},  
 where the GUT gauge symmetry is (partly) broken by 
 orbifold boundary conditions.  
In this paper, we consider a class of SO(10) models in 5D \cite{Raby}, 
 where SO(10) gauge symmetry is broken into the PS gauge group 
 and further symmetry breaking into the SM gauge group 
 is achieved by VEVs of Higgs multiplets on a brane. 
In particular, we concentrate on the recently proposed 
 simple SO(10) model \cite{F-O}. 
In this model, all matter and Higgs multiplets reside only 
 on a brane (PS brane) where the PS gauge symmetry is manifest, 
 so that low energy effective description of this model 
 is nothing but the PS model in 4D with a special set of 
 matter and Higgs multiplets. 
At energies higher than the compactification scale, 
 the Kaluza-Klein (KK) modes of the bulk SO(10) gauge multiplet 
 are involved in the particle contents and in fact, the gauge coupling 
 unification was shown to be successfully realized 
 by incorporating the KK mode threshold corrections 
 into the gauge coupling running \cite{F-O}. 
The unification scale ($M_{\rm GUT}$) and 
 the compactification scale ($M_c$) which was set to be the same 
 as the PS symmetry breaking scale ($v_{\rm PS}$) were found to be 
 $M_{\rm GUT}=4.6 \times 10^{17}$ GeV and 
 $M_c=v_{\rm PS}=1.2 \times 10^{16}$ GeV.

In this paper, we apply this SO(10) model to the inflationary scenario. 
The idea of inflation \cite{InflationRev} has been strongly favored 
 from the view points of not only providing the solutions 
 to the horizon and flatness problems of the standard big bang cosmology 
 but also recent precise cosmological observations on the 
 the cosmic microwave background radiation and the large 
 scale structure in the Universe. 
Therefore, it is an important task to construct a realistic 
 inflation model based on some well-motivated particle physics model. 
Among many proposed inflation models, the hybrid inflation 
 \cite{Linde} \cite{HIRev}
 is particularly  attractive because it can be adjusted to 
 a wide class of SUSY models \cite{Hybrid1}. 
Variants of hybrid inflation, in particular, 
 applicable to SUSY GUT models have been proposed: 
 The standard \cite{StdHI}, shifted \cite{ShiftedHI} 
 and smooth \cite{SmoothHI} hybrid inflation models. 
Some of these models are based on the SUSY PS model 
 with one singlet and Higgs multiplets 
 $({\bf 4},{\bf 1},{\bf 2}) \oplus (\overline{{\bf 4}},{\bf 1},{\bf 2})$ 
 whose VEVs break the PS symmetry to the SM one. 
Interestingly, except for the singlet field, the orbifold GUT model 
 of Ref.~\cite{F-O}, which we are interested in, has the same particle content. 
Therefore, the GUT model can naturally incorporate 
 the hybrid inflation in it. 
In the following, we consider the smooth hybrid inflation 
 \cite{SmoothHI} in this orbifold GUT framework.

This paper is organized as follows: 
In Section 2 we briefly review the model proposed in Ref.~\cite{F-O}. 
Application of this model to inflation is developed in Section 3 
 and the consistency of the model with current observations is shown. 
The last section is devoted for conclusions.

\section{Model Setup}
Here we briefly review the orbifold SO(10) GUT model 
 proposed in Ref.~\cite{F-O}. 
The model is described in 5D and 
 the 5th dimension is compactified 
 on the orbifold $S^1/{Z_2 \times Z_2^\prime}$. 
A circle $S^1$ with radius $R$ is divided by 
 a $Z_2$ orbifold transformation $y \to -y$ 
 ($y$ is the fifth dimensional coordinate $ 0 \leq y < 2 \pi R$)
 and this segment is further divided by a $Z_2^\prime$ transformation 
 $y^\prime \to -y^\prime $ with $y^\prime = y + \pi R/2$. 
There are two inequivalent orbifold fixed points at $y=0$ and $y=\pi R/2$. 
Under this orbifold compactification, a general bulk wave function 
 is classified with respect to its parities,  
 $P=\pm$ and $P^\prime=\pm$, under $Z_2$ and $Z_2^\prime$, respectively.

Assigning the parity ($P,P^\prime $) 
 the bulk SO(10) gauge multiplet as listed in Table~I, 
 only the PS gauge multiplet has zero-mode 
 and the bulk 5D N=1 SUSY SO(10) gauge symmetry is broken 
 to 4D N=1 SUSY PS gauge symmetry. 
Since all vector multiplets has wave functions  
 on the brane at $y=0$, SO(10) gauge symmetry is respected there, 
 while only the PS symmetry is on the brane at $y=\pi R/2$ (PS brane). 

\begin{table}[h]
\begin{center}
\begin{tabular}{|c|c|c|}
\hline
$(P,P')$ & bulk field & mass\\
\hline 
& & \\
$(+,+)$ & $V(15,1,1)$, $V(1,3,1)$, $V(1,1,3)$ & $\frac{2n}{R}$\\
& & \\
\hline
& & \\
$(+,-)$ &  $V(6,2,2)$ & $\frac{(2n+1)}{R}$ \\
& & \\
\hline
& & \\
$(-,+)$ &  $\Phi (6,2,2)$
& $\frac{(2n+1)}{R}$\\
& & \\
\hline
& & \\
$(-,-)$ & $\Phi (15,1,1)$, $\Phi (1,3,1)$, $\Phi (1,1,3)$ & $\frac{(2n+2)}{R}$ \\
& & \\
\hline
\end{tabular}
\end{center}
\caption{
 ($P,~P^\prime$) assignment and masses ($n \geq 0$) of fields 
 in the bulk SO(10) gauge multiplet $(V,~\Phi)$ 
 under the PS gauge group. 
$V$ and $\Phi$ are the vector multiplet and adjoint chiral 
 multiplet in terms of 4D N=1 SUSY theory. 
}
\label{t1}
\end{table}

We place the all matter and Higgs multiplets on the PS brane, 
 where only the PS symmetry is manifest 
 so that the particle contents are in the representation 
 under the PS gauge symmetry, not necessary to be 
 in SO(10) representation.   
For a different setup, see \cite{Raby}. 
The matter and Higgs in our model is listed in Table~2. 
For later conveniences, let us introduce the following notations: 
\bea
H_1&=&({\bf 1},{\bf 2},{\bf 2})_H, ~H_1^{\prime}=({\bf 1},{\bf 2},{\bf 2})'_H,
\nonumber \\
H_6&=&({\bf 6},{\bf 1},{\bf 1})_H, ~H_{15}=({\bf 15},{\bf 1},{\bf 1})_H,
\nonumber \\ 
H_L&=&({\bf 4},{\bf 2},{\bf 1})_H,
~\overline{H_L} =(\overline{{\bf 4}},{\bf 2},{\bf 1})_H,  
\nonumber \\
\phi&=&({\bf 4},{\bf 1},{\bf 2})_H,
~\bar{\phi}=(\overline{{\bf 4}},{\bf 1},{\bf 2})_H.
\eea

Superpotential relevant for fermion masses is given by%
\footnote{
For simplicity, we have introduced only minimal terms 
 necessary for reproducing observed fermion mass matrices. 
}
\bea
W_Y&=& Y_{1}^{ij} F_{Li} F_{Rj}^c H_1 
+\frac{Y_{15}^{ij}}{M_5} F_{Li} F_{Rj}^c 
 \left(H_1^{\prime} H_{15} \right) \nonumber\\ 
&+&\frac{Y_R^{ij}}{M_5} F_{Ri}^c F_{Rj}^c 
 \left(\phi \phi \right)  
 +\frac{Y_L^{ij}}{M_5} F_{Li}F_{Lj} 
 \left(\overline{H_L} \overline{H_L} \right), 
\label{Yukawa}
\eea 
where $M_5$ is the 5D Planck scale. 
The product, $H_1^{\prime} H_{15}$, effectively works 
 as $({\bf 15},{\bf 2},{\bf 2})_H$, 
 while $\phi \phi$ and $\overline{H_L}\overline{H_L}$ 
 effectively work as $({\bf 10},{\bf 1},{\bf 3})$ and 
 $(\overline{{\bf 10}},{\bf 3},{\bf 1})$, respectively, 
 and are responsible for the left- and the right-handed 
 Majorana neutrino masses. 
Providing VEVs for appropriate Higgs multiplets, 
 fermion mass matrices are obtained. 
There are a sufficient number of free parameters 
 to fit all the observed fermion masses and mixing angles. 

\begin{table}[h]
{\begin{center}
\begin{tabular}{|c|c|}
\hline
& brane at $y=\pi R/2$ \\ 
\hline
& \\
Matter Multiplets & $\psi_i=F_{Li} \oplus F_{Ri}^c \quad (i=1,2,3)$ \\
 & \\
\hline
 & \\
Higgs Multiplets & 
$({\bf 1},{\bf 2},{\bf 2})_H$,  
$({\bf 1},{\bf 2},{\bf 2})'_H$,
$({\bf 15},{\bf 1},{\bf 1})_H$,
$({\bf 6},{\bf 1},{\bf 1})_H$ \\  & 
$({\bf 4},{\bf 1},{\bf 2})_H$, 
$(\overline{{\bf 4}},{\bf 1},{\bf 2})_H$, 
$({\bf 4},{\bf 2},{\bf 1})_H$, 
$(\overline{{\bf 4}},{\bf 2},{\bf 1})_H$  \\
& \\
\hline
\end{tabular}
\end{center}}
\caption{
Particle contents on the PS brane. 
$F_{Li}$ and $F_{Ri}^c$ are matter multiplets 
 of i-th generation in $({\bf 4, 2, 1})$ and $({\bf \bar{4}, 1, 2})$ 
 representations, respectively. 
}
\end{table}

Suppose a Higgs superpotential which provides 
 the same VEVs ($v_{\rm PS}$) for Higgs multiplets to break the PS symmetry 
 and leaves only the particle contents of the minimal supersymmetric 
 Standard Model (MSSM) at low energies. 
In Ref.~\cite{F-O}, assuming $M_c=v_{\rm PS}$ and imposing 
 the left-right symmetry, the gauge coupling unification was examined. 
Analyzing the gauge coupling runnings in the MSSM, 
 $M_c=v_{\rm PS}$ is fixed as the scale 
 where the SU(2)$_L$ and SU(2)$_R$ gauge couplings coincide 
 with each other, which is found to be  
 $M_c=v_{\rm PS}=1.2 \times 10^{16}$ GeV. 
For the scale $\mu \geq M_c=v_{\rm PS}$,  
 we have only two independent gauge couplings, 
 SU(4)$_c$ and SU(2)$_L$ (or SU(2)$_R$) gauge couplings. 
After taking KK mode contributions into account, 
 it was shown that the gauge coupling is successfully unified at  
 $M_{\rm GUT}=4.6 \times 10^{17}$ GeV 
 (see Figure~1 from Ref.~\cite{F-O}). 
We assume that a more fundamental SO(10) GUT theory 
 takes place at $M_{GUT}$, 
 and it would be natural to assume $M_{GUT} \sim M_5$. 
In fact, the relation between 4D and 5D Planck scales, 
 $M_5^3/M_c \simeq M_P^2$ 
 ($M_P=2.44 \times 10^{18}$ GeV is the reduced Planck scale), 
 supports this assumption with $M_c=1.2 \times 10^{16}$ GeV. 
When we abandon the left-right symmetry, 
 there is more freedom for the gauge coupling unification 
 with two independent parameters $v_{PS}$ and $M_c$.

Our model gives a large $v_{PS}$ relative to 
 other 5D orbifold SO(10) models \cite{Raby}. 
The high value of $v_{PS}$ or $M_c$ is advantageous 
 for dangerous proton decay due to dimension six operators. 
 From Eq.~(\ref{Yukawa}), the right-handed neutrino mass scale 
 is given by $M_R \sim Y_R v_{PS}^2/M_5 \sim Y_R v_{PS}^2/M_{\rm GUT}$. 
The scale $M_R = {\cal O}(10^{14}~\mbox{GeV})$ 
 preferable for the seesaw mechanism can be obtained 
 by a mild tuning of the Yukawa coupling $Y_R \sim 0.1$.

In the next section, we show that this model is well fitted 
 to the smooth hybrid inflation model and the model predictions 
 are compatible with the cosmological observations 
 like the power spectrum of the curvature perturbations, 
 the scalar spectral index, the ratio of scalar-to-tensor 
 fluctuations and so on.

\section{Smooth hybrid inflation}
We consider the smooth hybrid inflation model \cite{SmoothHI} 
 in the context of the orbifold GUT model 
 discussed in the previous section. 
For this purpose, we introduce a singlet chiral superfield $S$.  
Needless to say, this singlet field causes 
 no change for the gauge coupling unification. 
Let us consider the superpotential 
 for the smooth hybrid inflation \cite{SmoothHI}%
\footnote{
The renormalizable term, $S (\bar{\phi} \phi)$, 
 can be forbidden by introducing a discrete symmetry \cite{SmoothHI}, 
 for example, $\phi \to -\phi$ and $\bar{\phi} \to \bar{\phi}$. 
}, 
\bea
 W= S \left( -\mu^2+\frac{(\bar{\phi} \phi)^2}{M^2} \right), 
\label{smooth}  
\eea
where we have omitted possible ${\cal O}(1)$ coefficients. 
SUSY vacuum conditions lead to non-zero VEVs for 
 $\langle \phi \rangle = \langle \bar{\phi} \rangle = \sqrt{\mu M}$, 
 by which the PS symmetry is broken down to the SM one, and thus  
\bea 
  v_{\rm PS} = \sqrt{\mu M}.   
\eea 
In the following analysis of inflation, we treat $M$ as a
 free parameter with $v_{\rm PS} = \sqrt{\mu M}=1.2 \times 10^{16}$ GeV    
 fixed by the analysis of gauge coupling unification. 
Since $M$ is involved in the non-renormalizable term, 
 it is theoretically natural that 
 $M \sim M_5 \sim M_{\rm GUT}=4.6 \times 10^{17}$ GeV. 
Nevertheless this condition is independent of the inflation scenario, 
 we will find from the following analysis that 
 $M \sim M_{\rm GUT}$ is in fact consistent with the cosmological
 observations.
Therefore, our GUT model is suitable for the inflation models 
 with the parameters $v_{\rm PS}$ and $M_{\rm GUT}$ 
 fixed by the analysis for the gauge coupling unification. 

There are many possibility of the other parts of Higgs 
 superpotential involving other Higgs multiplets in Table~2. 
See, for example, Ref.~\cite{F-O} and also 
 the Higgs superpotential in Ref.~\cite{ShiftedHI}. 
It would be worth mentioning that the Higgs superpotential 
 includes a term $W \supset H_6 (\phi^2+\bar{\phi}^2)$ 
 through which all color triplets in $\phi$ and $\bar{\phi}$ become heavy. 
Only the superpotential of Eq.~(\ref{smooth}) is relevant for inflation.

Now we examine the smooth hybrid inflation scenario. 
The scalar potential from  Eq.~(\ref{smooth}) is given by%
\footnote{
Although supergravity effects from Kahler potential 
 can play an important role \cite{sugra}, 
 we do not consider effects form supergravity in this paper, 
 assuming a special Kahler potential for the inflaton field. 
}
\bea
 V= \left| -\mu^2+\frac{(\bar{\phi} \phi)^2}{M^2}\right|^2
 +4S^2 \frac{|\phi|^2 |\bar{\phi}|^2}{M^4} 
 \left(|\phi|^2 + |\bar{\phi}|^2\right). 
\label{vsmooth}
\eea
Considering the D-flatness condition, we normalize
\bea
 |\phi|=|\bar{\phi}|=\frac{\chi}{2},~~|S|=\frac{\sigma}{\sqrt{2}} 
\eea
and then V becomes
\bea
 V=\left( \mu^2-\frac{\chi^4}{16M^2} \right)^2 
 +\frac{\chi^6\sigma^2}{16M^4}. 
\eea
For a fixed $\sigma$, $V$ has a minimum at 
\bea
 \chi^2 = -6 \sigma^2 + \sqrt{36 \sigma^4 + 16 v_{\rm PS}^4}
        \simeq \frac{4}{3} \frac{v_{\rm PS}^4}{\sigma^2}, 
\eea
 where we have used an approximation for 
 $\sigma^2 \gg v_{PS}^2$ (satisfied during inflation) 
 in the last expression. 
The inflation trajectory is along this minimum and 
 we obtain the potential along this path 
\bea
  V \simeq \mu^4 
  \left( 1 - \frac{2 v_{\rm PS}^4}{27 \sigma^4} \right) .  
\eea
Note that the PS symmetry is spontaneously broken 
 anywhere on the inflation trajectory  and thus, 
 no topological defects such as strings, monopoles, 
 or domain walls are produced at the end of inflation
\cite{SmoothHI}.

Accordingly, the slow-roll parameters ($\epsilon$, $\eta$) 
 and the parameter ($\xi^2$), which enters the running of the 
 spectral index, are defined as \cite{InflationRev} 
\bea
 \epsilon &=& \frac{M_P^2}{2} \left( \frac{V'}{V} \right)^2 
 \simeq \frac{32 M_P^2 v_{\rm PS}^8}{729 \sigma^{10}} 
 \simeq - \frac{4 v_{\rm PS}^4}{135 \sigma^4} \eta \nonumber \\
 \eta &=& M_P^2 \left( \frac{V''}{V} \right) = 
  - \frac{40 M_P^2 v_{\rm PS}^4}{27 \sigma^6}  \nonumber \\
 \xi^2 &=& M_P^4 \left( \frac{V' V'''}{V^2} \right) 
   \simeq 
   \frac{640 M_P^4 v_{\rm PS}^8}{243 \sigma^{12}},  
 \label{slow-roll}
\eea
where the prime denotes derivative with respect to $\sigma$. 
The slow-roll approximation is valid if the conditions,  
 $\epsilon$ and $|\eta| \ll 1$, hold. 
In this case, 
 the spectral index ($n_{\rm s}$), 
 the ratio of tensor-to-scalar fluctuations ($r$) 
 and the running of the spectral index ($\alpha_{\rm s}$) 
 are given by 
\bea 
 n_{\rm s} &\simeq &1-6 \epsilon + 2 \eta,  \nonumber \\
 r &\simeq &16 \epsilon,  \nonumber \\
 \alpha_{\rm s} &=& \frac{d n_{\rm s}}{d \ln k}
  \simeq 16 \epsilon \eta -24 \epsilon^2 -2 \xi^2. 
\label{spectral}
\eea

The number of e-folds $N_k$ after the comoving scale $\ell_0=2 \pi/k_0$ 
 has crosses the horizon is given by 
\bea
N_k = \frac{1}{M_P^2} \int^{\sigma_k}_{\sigma_f} 
  d\sigma \frac{V}{V'} 
  \simeq \frac{9}{16 M_P^2 v_{\rm PS}^4} 
  \left( \sigma_k^6 - \sigma_f^6  \right),  
 \label{e-folds1}
\eea
where $\sigma_k$ is the value of the inflaton field 
 when the scale corresponding to $k_0$ 
 exits the horizon, and $\sigma_f$ is the value of the inflaton field 
 when the inflation ends, which is determined by $|\eta|=1$ 
 so that 
\bea
\sigma_f^6 &=& \frac{40}{27} M_P^2 v_{\rm PS}^4.  
\eea
For $\sigma_k^6 \gg \sigma_f^6$, several formulas given above 
 are reduced into simpler forms, for example, 
\bea 
 N_k \simeq - \frac{5}{6 \eta_k}, ~~
 n_s \simeq 1- \frac{5}{3 N_k}, ~~ 
 \alpha_{\rm s} \simeq -\frac{5}{3 N_k^2} .  
\eea 
The number of e-folds $N_k$ required for solving the horizon  
 and flatness problems of the standard big bang cosmology, 
 for $k_0=0.002$ Mpc$^{-1}$, is given by \cite{InflationRev}
\bea 
 N_k \simeq 51.4 
 + \frac{2}{3} \ln \left( \frac{V(\sigma_k)^{1/4}}{10^{15}~{\rm GeV}} \right)
 + \frac{1}{3} \ln \left( \frac{T_{\rm rh}}{10^{7}~{\rm GeV}} \right), 
\label{e-folds2}
\eea 
 where we have assumed a standard thermal history, 
 and $T_{\rm rh}$ is the reheating temperature after inflation. 
If gravitino has mass around 100 GeV as in the gravity mediated 
 SUSY breaking, the reheating temperature is severely constraint 
 (gravitino problem) 
 in order for the gravitino decay products not to destroy 
 the light elements successfully synthesized during 
 big bang nucleosynthesis \cite{gravitino}, 
\bea
 T_{\rm rh} \leq  10^6-10^7 \textrm{GeV}.  
\eea
The power spectrum of the primordial curvature perturbation 
 at the scale $k_0$ is given by 
\bea 
 {\cal{P_R}}^{1/2} \simeq 
\frac{1}{2 \sqrt{3} \pi M_P^3} \frac{V^{3/2}}{|V'|} 
\simeq 
 \frac{27}{16 \sqrt{3} \pi } \frac{\sigma_k^5}{M_P^3 M^2}.  
 \label{power}
\eea
This should satisfy the observed value by the WMAP \cite{WMAP}, 
 ${\cal{P_R}} \simeq  2.457 \times 10^{-9}$.

Now we solve Eqs.~(\ref{e-folds1}), (\ref{e-folds2}) 
 and (\ref{power}) and fix the model parameters 
 involving the inflation scenario. 
In our analysis, we have three independent free parameters,  
 $M$, $\sigma_k$ and $T_{\rm rh}$, 
 with the fixed $v_{\rm PS}=1.2 \times 10^{16}$ GeV. 
We solve the equations for a given $T_{\rm rh}$ 
 in the range 1 MeV $\leq T_{\rm rh} \leq 10^{10}$ GeV 
 and find $M$ and $\sigma_k$. 
Figure~2 shows the results for $M$ and $\sigma_k$ 
 as a function of $T_{\rm rh}$. 
We can check that $\sigma_k^6 \gg \sigma_f^6$ 
 and the slow-roll conditions are satisfied. 
The number of e-folds is depicted in Figure~3. 
Using these outputs and Eq.~(\ref{slow-roll}), 
 we evaluate the spectral index (Figure~4), 
 the tensor-to-scalar ratio and the running of the spectral index: 
\bea 
   0.963 \leq  & n_{\rm s} & \leq 0.968,   \nonumber \\ 
  4.0 \times 10^{-7} \geq  & r &   \geq  3.1 \times 10^{-7} ,
  \nonumber \\ 
  -8.4 \times 10^{-4} \leq & \alpha_{\rm s} & \leq -6.1 \times 10^{-4} 
\eea  
 for 1 MeV $\leq T_{\rm rh} \leq 10^7$ GeV. 
The tensor-to-scalar ratio and the running of the spectral index 
 are negligibly small. 
These results are consistent with the WMAP 5-year data \cite{WMAP}: 
 $n_{\rm s} = 0.960^{+0.014}_{-0.013}$, 
 $ r <  0.2$ (95\% CL) 
 and $ \alpha_{\rm s} = -0.032^{+0.021}_{-0.020}$ (68\% CL) 
 (consistent with zero in 95\% CL).

In general, we do not need to impose either the left-right symmetry 
 or $M_c=v_{\rm PS}$ on the model. 
In this case, $v_{\rm PS}$ can be varied \cite{Raby}, 
 and we repeat the same analysis for this general case 
 with $v_{\rm PS}$ as a free parameter. 
Fixing, for example, $T_{\rm rh}=10^7$ GeV, 
 $M$ can be obtained as a function of $v_{\rm PS}$ 
 as shown in Figure~5. 
Since $M$ appears in the non-renormalizable term, 
 it would be natural to identify $M$ as an effective cutoff. 
Thus, the theoretical consistency leads to the condition, 
 $v_{\rm PS} \leq M \leq M_P$, from which we obtain 
 $1.9 \times 10^{14}$ GeV $\leq v_{\rm PS} \leq 5.6 \times 10^{16}$ GeV. 
The number of e-folds and the tensor-to-scalar ratio 
 are shown in Figure~6 and 7, respectively, 
 as a function of $v_{\rm PS}$. 
The other outputs, the spectral index and its running, 
 are found as 
\bea  
  0.967 \leq &n_{\rm s}& \leq 0.968,  \nonumber \\   
  -6.4 \times 10^{-4} \leq  &\alpha_{\rm s}&  
  \leq -6.1 \times 10^{-4}.  
\eea 
These are consistent with the WMAP data.

\section{Conclusions} 
We have discussed the smooth hybrid inflation scenario 
 in the context of the SO(10) GUT model in Ref.~\cite{F-O}. 
We have shown that the model can be naturally applicable 
 to the inflation model by introducing only one singlet 
 chiral multiplet, keeping the structure of the gauge coupling 
 unification unchanged. 
The analysis for the running gauge coupling fixes 
 the model parameters as $v_{\rm PS}= 1.2 \times 10^{16}$ GeV 
 and $M_{\rm GUT} = 4.6 \times 10^{17}$ GeV. 
It is very interesting that these parameters are determined 
 independently of the cosmological considerations, 
 nevertheless the inflation scenario with these parameters 
 fits very well with the current cosmological observations, 
 as we have shown.

\section*{Acknowledgments}
The works of T.F. and N.O. are supported in part by 
 the Grant-in-Aid for Scientific Research from the Ministry 
 of Education, Science and Culture of Japan 
 (\#20540282 and \#18740170, respectively). 



\newpage

\begin{figure}[t]
\begin{center}
\includegraphics[scale=1.2]{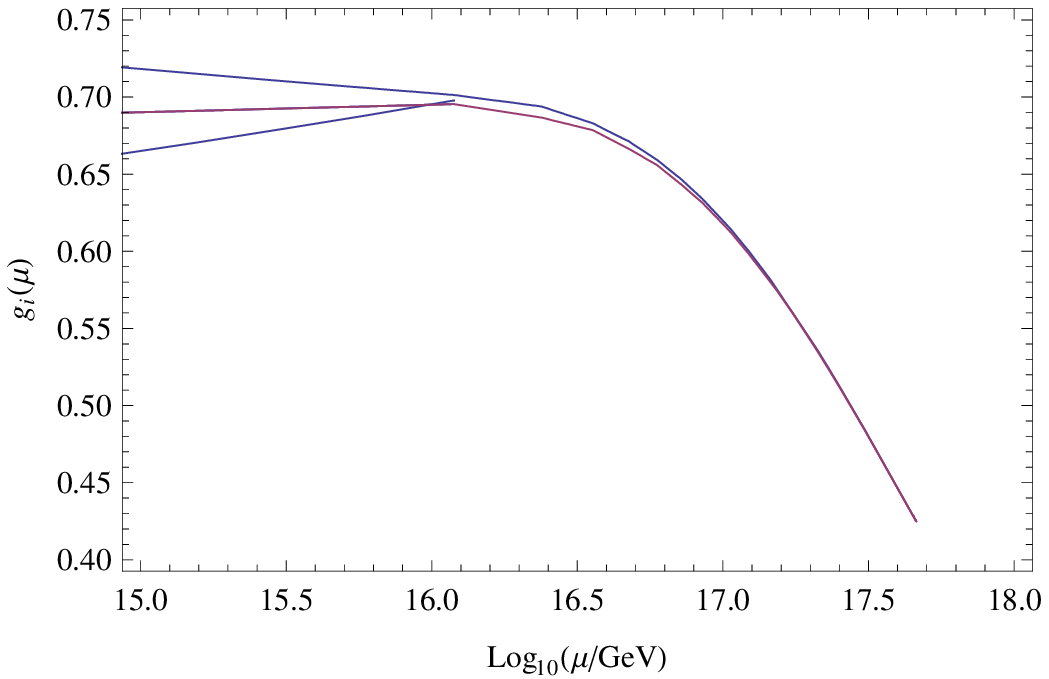}
\caption{
Gauge coupling unification in the left-right symmetric case, 
 taken from Ref.~\cite{F-O}. 
Each line from top to bottom corresponds to 
 $g_3$, $g_2$ and $g_1$ for $ \mu < M_c=v_{\rm PS}$, 
 while  $g_3=g_4$ and $g_2=g_{2R}$ for $ \mu > M_c=v_{\rm PS}$. 
}
\end{center}
\end{figure}
\begin{figure}[t]
\begin{center}
\includegraphics[scale=1.2]{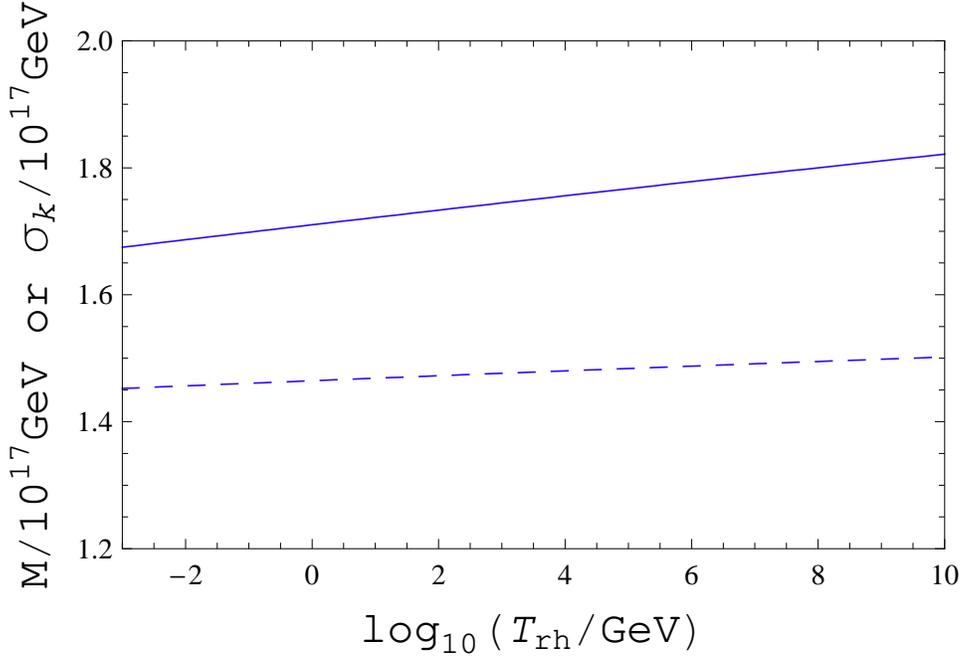}
\caption{
$M$ and $\sigma_k$ as a function of $T_{\rm rh}$. 
The solid and dashed lines correspond to 
 $M$ and $\sigma_k$, respectively. 
}
\end{center}
\end{figure}
\begin{figure}[t]
\begin{center}
\includegraphics[scale=1.2]{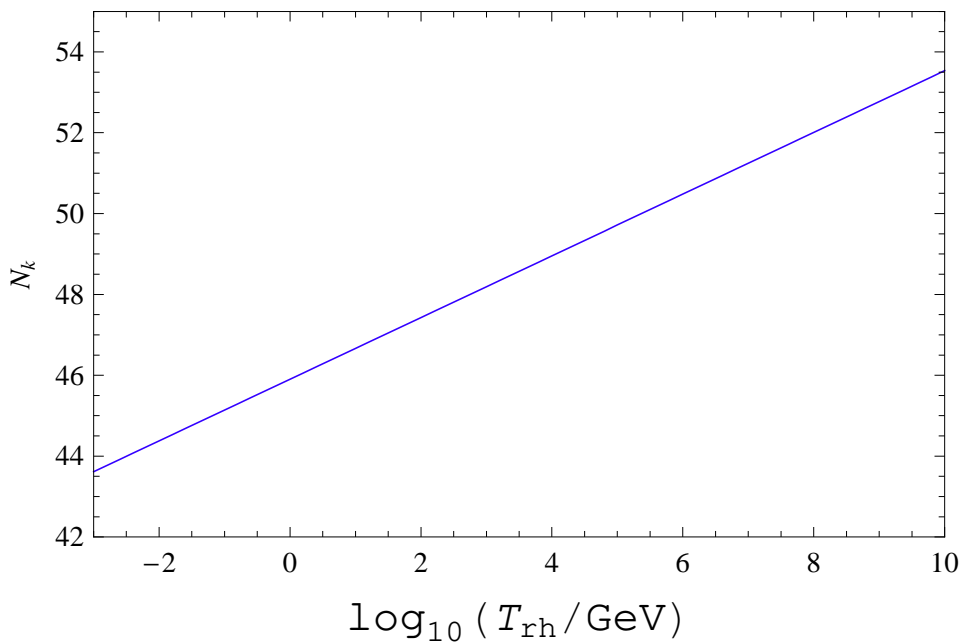}
\caption{ 
The number of e-folds versus $T_{\rm rh}$. 
}
\end{center}
\end{figure}
\begin{figure}[t]
\begin{center}
\includegraphics[scale=1.2]{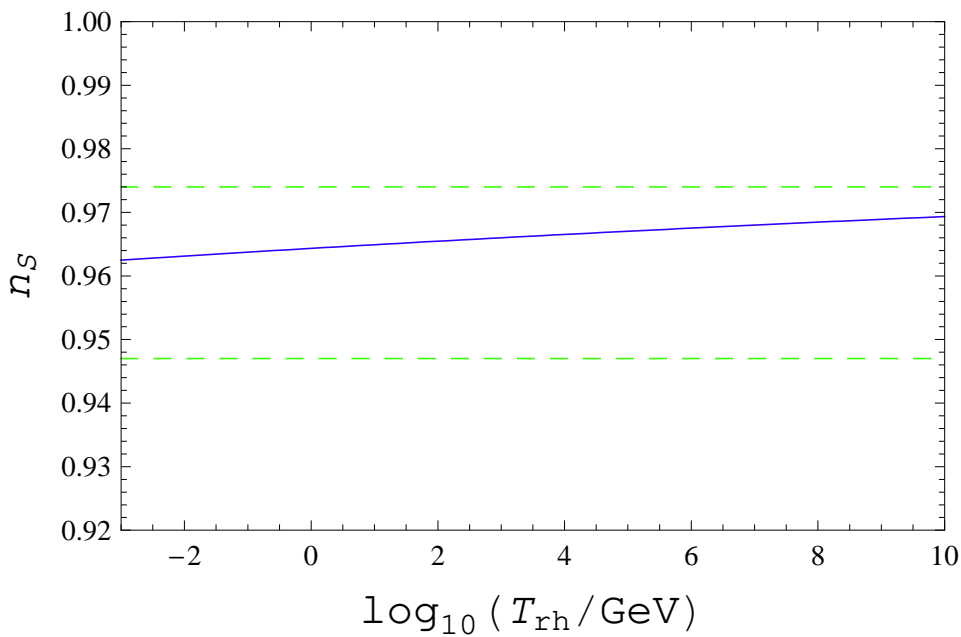}
\caption{ 
The spectral index as a function of $T_{\rm rh}$. 
The dashed lines correspond to the WMAP 5-year data, 
 $n_{\rm s} = 0.960^{+0.014}_{-0.013}$. 
}
\end{center}
\end{figure}
\begin{figure}[t]
\begin{center}
\includegraphics[scale=1.2]{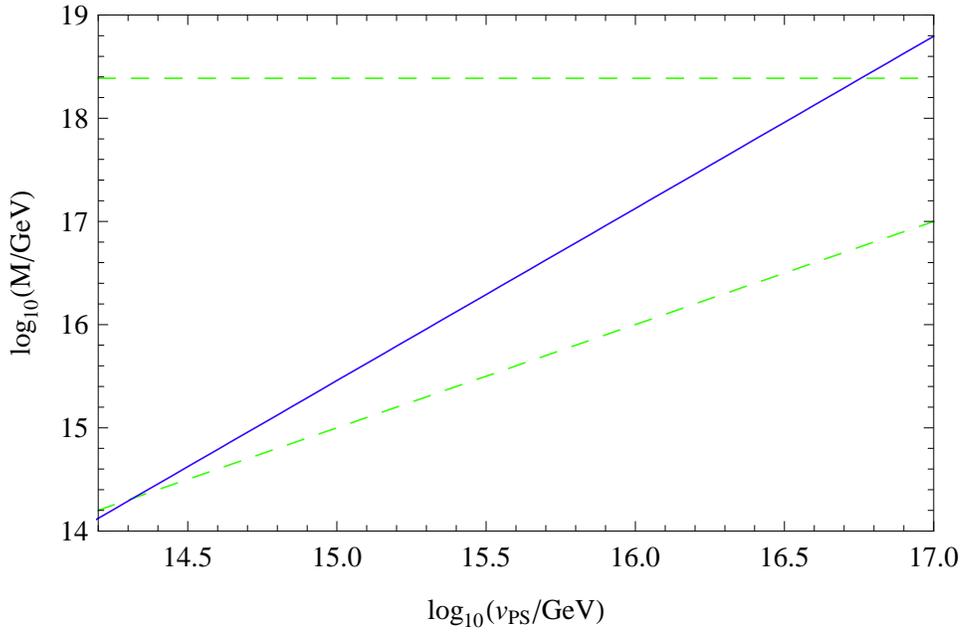}
\caption{ 
$M$ as a function of $v_{\rm PS}$ (solid line). 
The lower and upper dashed lines specified 
 the theoretical consist region for $M$, 
 $ v_{\rm PS} \leq M \leq M_P$. 
}
\end{center}
\end{figure}
\begin{figure}[t]
\begin{center}
\includegraphics[scale=1.2]{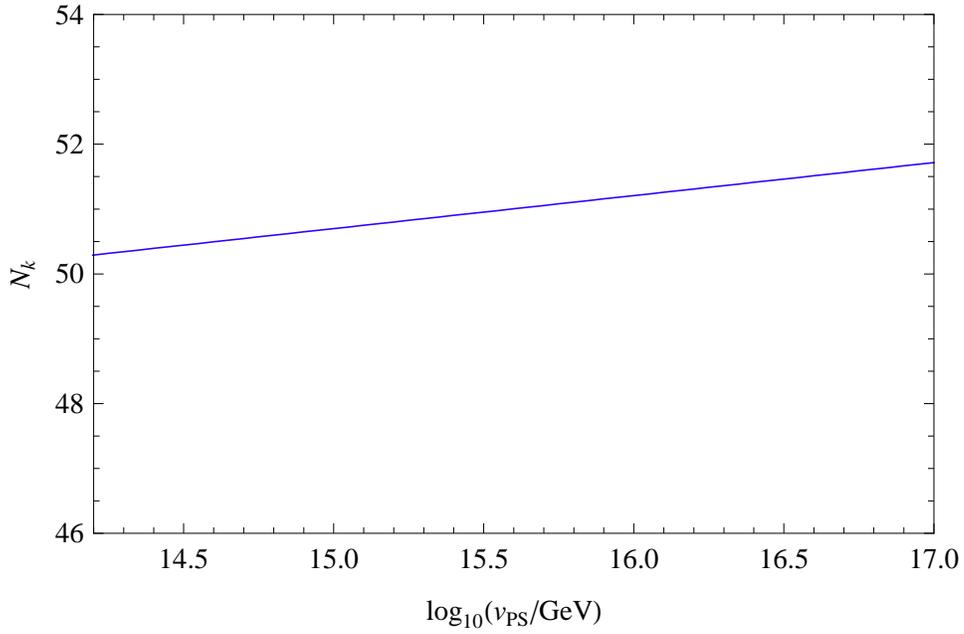}
\caption{ 
The number of e-folds versus $v_{\rm PS}$. 
}
\end{center}
\end{figure}
\begin{figure}[t]
\begin{center}
\includegraphics[scale=1.2]{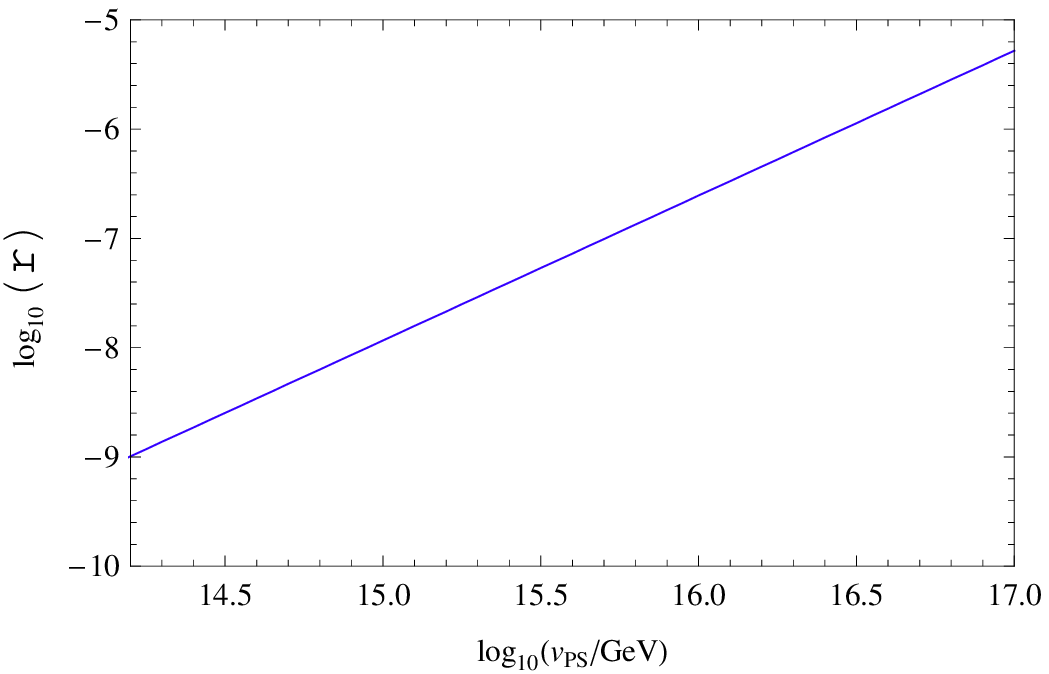}
\caption{ 
The tensor-to-scalar ratio versus $v_{\rm PS}$. 
}
\end{center}
\end{figure}


\begin{thebibliography}{99}

\bibitem{Babu}
  K.~S.~Babu and R.~N.~Mohapatra,
  Phys.\ Rev.\ Lett.\  {\bf 70}, 2845 (1993). 

\bibitem{Fukuyama1}
  K.~Matsuda, Y.~Koide and T.~Fukuyama,
  Phys.\ Rev.\ D {\bf 64}, 053015 (2001); 
  K.~Matsuda, Y.~Koide, T.~Fukuyama and H.~Nishiura,
  Phys.\ Rev.\ D {\bf 65}, 033008 (2002)
  [Erratum-ibid.\ D {\bf 65}, 079904 (2002)]. 

\bibitem{Fukuyama2}
  T.~Fukuyama and N.~Okada,
  JHEP {\bf 0211}, 011 (2002). 

\bibitem{Eguchi:2002dm}
  K.~Eguchi {\it et al.}  [KamLAND Collaboration],
  Phys.\ Rev.\ Lett.\  {\bf 90}, 021802 (2003). 

\bibitem{Fukuyama:2004xs}
  T.~Fukuyama, A.~Ilakovac, T.~Kikuchi, S.~Meljanac and N.~Okada,
  Eur.\ Phys.\ J.\ C {\bf 42}, 191 (2005); 
{\it ibid.},
J.\ Math.\ Phys.\  {\bf 46}, 033505 (2005); 
%
Phys.\ Rev.\ D {\bf 72}, 051701 (2005). 


\bibitem{Bajc:2004xe}
  B.~Bajc, A.~Melfo, G.~Senjanovi\'c and F.~Vissani,
  Phys.\ Rev.\ D {\bf 70}, 035007 (2004). 

\bibitem{Bertolini}
  S.~Bertolini, T.~Schwetz and M.~Malinsky,
  Phys.\ Rev.\ D {\bf 73}, 115012 (2006). 


\bibitem{F-K-O}
 T.~Fukuyama, T.~Kikuchi and N.~Okada,
  Phys.\ Rev.\  D {\bf 75}, 075020 (2007);  
 R.~N.~Mohapatra, N.~Okada and H.~B.~Yu,
  Phys.\ Rev.\  D {\bf 76} 015013 (2007). 

\bibitem{RS}
 L.~Randall and R.~Sundrum,
 Phys.\ Rev.\ Lett.\  {\bf 83}, 3370 (1999). 


\bibitem{Kawamura}
Y.~Kawamura, Prog.\ Theor.\ Phys.\ {\bf 105}, 999 (2001); . 
G.~Altarelli and F.~Feruglio,
 Phys.\ Lett.\  B {\bf 511}, 257 (2001);  
A.~B.~Kobakhidze,
 Phys.\ Lett.\  B {\bf 514}, 131 (2001); .
L.~Hall and Y.~Nomura, 
 Phys.\ Rev.\ D {\bf 64} 055003 (2001); 
A.~Hebecker and J.~March-Russell,
 Nucl.\ Phys.\  B {\bf 613}, 3 (2001). 


\bibitem{Raby}
R.~ Dermisek and A.~Mafi, 
 Phys.\ Rev.\ D {\bf 65}, 055002 (2002); 
%
H.~D.~Kim and S.~Raby, 
 JHEP {\bf 0301} 056 (2003); 
%
S.~M.~Barr and I.~Dorsner,
 Phys.\ Rev.\  D {\bf 66}, 065013 (2002); 
%
I.~Dorsner,
 Phys.\ Rev.\  D {\bf 69}, 056003 (2004);  
%
B.~Kyae, C.~A.~Lee and Q.~Shafi,
 Nucl.\ Phys.\  B {\bf 683}, 105 (2004); 
%
M.~L.~Alciati and Y.~Lin,
 JHEP {\bf 0509} (2005) 061; 
%
M.~L.~Alciati, F.~Feruglio, Y.~Lin and A.~Varagnolo,
 JHEP {\bf 0611} (2006) 039. 


\bibitem{F-O}
T.~Fukuyama and N.~Okada,
  arXiv:0803.1758 [hep-ph], to be published 
  in Phys.\ Rev.\ D. 


\bibitem{Linde}
A.~D.~Linde,
 Phys.\ Rev.\  D {\bf 49}, 748 (1994). 


\bibitem{HIRev}
 For a review, see, for example, 
 G.~Lazarides,
  arXiv:hep-ph/0011130; 
%
 C.~Pallis,
  arXiv:0710.3074 [hep-ph], 
 and references therein. 


\bibitem{Hybrid1}
 E.~J.~Copeland, A.~R.~Liddle, D.~H.~Lyth, E.~D.~Stewart and D.~Wands,
  Phys.\ Rev.\  D {\bf 49}, 6410 (1994). 


\bibitem{StdHI} 
 G.~R.~Dvali, Q.~Shafi and R.~K.~Schaefer,
  Phys.\ Rev.\ Lett.\  {\bf 73}, 1886 (1994); 
%
 G.~Lazarides, R.~K.~Schaefer and Q.~Shafi,
  Phys.\ Rev.\  D {\bf 56}, 1324 (1997). 


\bibitem{ShiftedHI} 
 R.~Jeannerot, S.~Khalil, G.~Lazarides and Q.~Shafi,
  JHEP {\bf 0010}, 012 (2000);  
 R.~Jeannerot, S.~Khalil and G.~Lazarides,
  JHEP {\bf 0207}, 069 (2002). 


\bibitem{SmoothHI}
G.~Lazarides and C.~Panagiotakopoulos,
  Phys.\ Rev.\  D {\bf 52}, 559 (1995); 
%
G.~Lazarides, C.~Panagiotakopoulos and N.~D.~Vlachos,
  Phys.\ Rev.\  D {\bf 54}, 1369 (1996); 
%
G.~Lazarides and A.~Vamvasakis,
  Phys.\ Rev.\  D {\bf 76}, 083507 (2007). 

\bibitem{sugra} 
A.~D.~Linde and A.~Riotto,
  Phys.\ Rev.\  D {\bf 56}, 1841 (1997); 
%
V.~N.~Senoguz and Q.~Shafi,
  Phys.\ Lett.\  B {\bf 567}, 79 (2003). 


\bibitem{InflationRev} 
For a review, see, for example, 
 D.~H.~Lyth and A.~Riotto,
  Phys.\ Rept.\  {\bf 314}, 1 (1999), and references therein; 
 Liddle A R and Lyth D H, {\it Cosmological 
 Inflation and Large-Scale Structure}, Cambridge University Press (2000).
%
  G.~Lazarides,
  Lect.\ Notes Phys.\  {\bf 592}, 351 (2002); 
  J.\ Phys.\ Conf.\ Ser.\  {\bf 53}, 528 (2006). 

\bibitem{gravitino} 
For recent analysis, see, for example, 
R.~H.~Cyburt, J.~R.~Ellis, B.~D.~Fields and K.~A.~Olive,
 Phys.\ Rev.\ D {\bf 67}, 103521 (2003); 
M.~Kawasaki, K.~Kohri and T.~Moroi,
 Phys.\ Lett.\  B {\bf 625}, 7 (2005). 

\bibitem{WMAP}
G.~Hinshaw {\it et al.}  [WMAP Collaboration],
  arXiv:0803.0732 [astro-ph].


\end{thebibliography}
\end{document}